\documentclass{ar-1col}
\usepackage[numbers]{natbib}
\usepackage{url}

\jname{Xxxx. Xxx. Xxx. Xxx.}
\jvol{AA}
\jyear{2021}
\doi{10.1146/((please add article doi))}

\begin{document}

% Page header
\markboth{Chen, I.Y., Pierson, E., Rose, S., Joshi, S., Ferryman, K., Ghassemi, M.}{Ethical Machine Learning in Health Care}

% Title
\title{Ethical Machine Learning in Health Care}

%Authors, affiliations address.
\author{Irene Y. Chen,$^1$ Emma Pierson,$^2$ Sherri Rose,$^3$ Shalmali Joshi,$^4$ Kadija Ferryman,$^5$\\and Marzyeh Ghassemi$^{4,6}$
\affil{$^1$Electrical Engineering and Computer Science, Massachusetts Institute of Technology, Cambridge, MA, 02139, USA; email: iychen@mit.edu}
\affil{$^2$Microsoft Research, Cambridge, MA, 02143, USA}
\affil{$^3$Center for Health Policy and Center for Primary Care and Outcomes Research, Stanford University, Stanford, CA, 94305, USA}
\affil{$^4$Vector Institute, Toronto, ON, Canada}
\affil{$^5$Department of Technology, Culture, and Society, Tandon School of Engineering, New York University, Brooklyn, NY, 11201, USA}
\affil{$^6$Department of Computer Science, University of Toronto, Toronto, ON, Canada}}

%Abstract (150 words)
\begin{abstract}
The use of machine learning (ML) in health care raises numerous ethical concerns, especially as models can amplify existing health inequities. Here, we outline ethical considerations for equitable ML in the advancement of health care. Specifically, we frame ethics of ML in health care through the lens of social justice. We describe ongoing efforts and outline challenges in a proposed pipeline of ethical ML in health, ranging from problem selection to post-deployment considerations. We close by summarizing recommendations to address these challenges.
\end{abstract}

%Keywords, up to 6
\begin{keywords}
machine learning, bias, ethics, health, health care, health disparities
\end{keywords}
\maketitle

%Table of Contents
\tableofcontents

% Heading 1
\section{INTRODUCTION}
\label{chap29:sec1}

As machine learning (ML) models proliferate into many aspects of our lives, there is growing concern regarding their ability to inflict harm. In medicine, excitement about human-level performance~\cite{topol2019high} of ML for health is balanced against ethical concerns, such as the potential for these tools to exacerbate existing health disparities~\cite{ferryman_winn_2016,wiens2019no,ghassemi2020review,ghassemi2019practical}. For instance, recent work has demonstrated that state-of-the-art clinical prediction models underperform on women, ethnic and racial minorities, and those with public insurance~\cite{chen2019can}. Other research has shown that when popular contextual language models are trained on scientific articles, they complete clinical note templates to recommend ``hospitals'' for violent white patients and ``prison'' for violent Black patients~\cite{zhang2020hurtful}. Even more worrisome, health care models designed to optimize referrals to long-term care-management programs for millions of patients were found to exclude Black patients with similar health conditions compared to white patients from care management programs~\cite{obermeyer2019dissecting}.

A growing body of literature wrestles with the social implications of machine learning and technology. Some of this work, referred to as critical data studies, is from a social science perspective~\cite{boyd2012critical,dalton2016critical}, whereas other work leads with a technical and computer science perspective~\cite{zliobaite2015survey,barocas2017fairness,corbett2018measure}. While there is scholarship addressing social implications and algorithmic fairness in general, there has been less work at the intersection of health, ML, and fairness~\cite{chen2018my,rajkomar2018ensuring,ustun2019fairness}, despite the potential life-or-death impacts of those models~\cite{obermeyer2019dissecting, benjamin2019assessing}. 

\begin{marginnote}[]
\entry{Machine learning (ML)}{The study of computer algorithms that improve automatically through experience}
\entry{ML model}{An algorithm that has been trained on data for a specific use case}
\entry{Algorithm}{A finite sequence of well-defined instructions used to solve a class of problems}
\end{marginnote}

While researchers looking to develop ethical ML models for health can begin by drawing on bioethics principles~\cite{veatch2019basics,vayena2018machine}, these principles are designed to inform clinical care practices. How these principles could inform the ML model development pipeline remains understudied. 
\begin{marginnote}[]
\entry{Ethical pipeline}{The model development process and the corresponding ethical considerations}
\entry{Bioethics}{The study of ethical issues emerging from advances in biology and medicine}
\entry{Justice}{The principle that obligates equitably distributed benefits, risks, costs, and resources}
\entry{Beneficence}{The principle that requires that care be provided with the intent of doing good for the patient involved}
\entry{Non-maleficence}{The principle that forbids harm or injury to the patient, either through acts of commission or omission}
\end{marginnote}
We note that there has been significant work on other important ethical issues that relate to ML and health, including reviews of consent and privacy~\cite{kaye2012tension}, which we do not address here. Instead, we focus on equity in ML models that operate on health data. We focus primarily on differences between groups induced by, or related to, the model development pipeline, drawing on both the bioethics principle of justice, and the established social justice centering of public health ethics~\cite{powers2006social}. Unjust differences in quality and outcomes of health care between groups often reflect existing societal disparities for disadvantaged groups. We consider other bioethics principles such as beneficence and non-maleficence, but focus them primarily on groups of patients rather than on individuals.

\begin{figure}
    \centering
    \includegraphics[width=0.95 \textwidth]{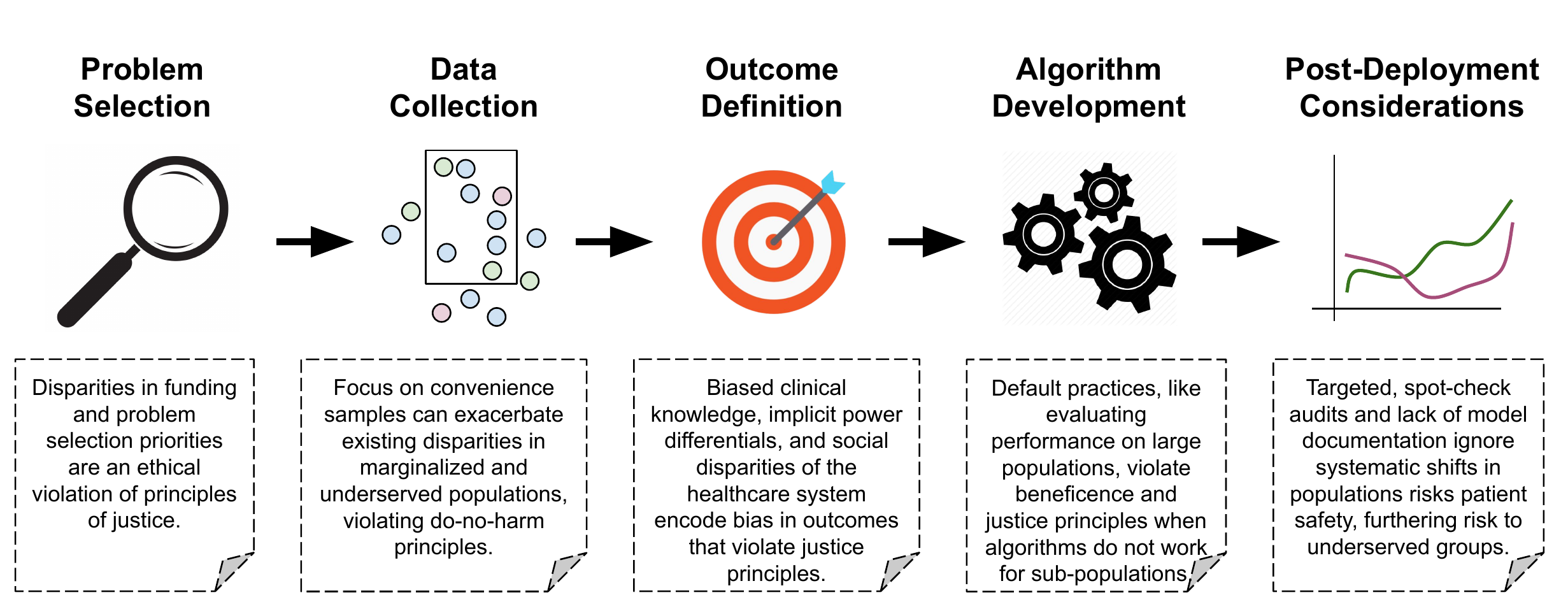}
    \caption{We motivate the five steps in the ethical pipeline for health care model development.  Each stage contains considerations for ML where ignoring technical challenges violate the bioethical principle of justice, either by exacerbating existing social injustices or by creating the potential for new injustices between groups. Although this review's ethical focus is on social justice, the challenges that we highlight may also violate ethical principles such as justice and beneficence. We highlight a few in this illustration.}
    \label{fig:pipeline}
\end{figure}

We organize this review by describing the ethical considerations that arise at each step of the pipeline during model development for ML in health (\textbf{Figure~\ref{fig:pipeline}}), from research funding to post-deployment. Here we motivate the ethical considerations in the pipeline with a case study of Black mothers in the United States, who die in childbirth at a rate three times higher than white women~\cite{berg1996pregnancy}. This inequity is unjust because it connects to a history of reproductive injustices faced by Black women in the United States, from gynecological experimentation on enslaved women to forced sterilizations~\cite{roberts1999killing,berry2017price}. 
\begin{enumerate}
\item This disparity occurs in part during \textbf{problem selection} because maternal mortality is an understudied problem~\cite{fisk2009systematic}.
\item Even after accounting for problem selection, \textbf{data collection} from hospitals may differ in quality and quantity. For example, 75\% of Black women give birth at hospitals that serve predominantly Black patients~\cite{howell2016black} but Black-serving hospitals have higher rates of maternal complications than other hospitals~\cite{creanga2014performance}.  
\item Once data are collected, the choice of \textbf{outcome definition} can obscure underlying issues, e.g., differences in clinical practice. General model outcome definitions for maternal health complications might overlook conditions specific to Black mothers, e.g., fibroids~\cite{eltoukhi2014health}.
\item During \textbf{algorithm development}, models may not be able to account for the confounding presence of societal bias. Black mothers in the wealthiest neighborhoods in Brooklyn, New York have worse outcomes than white, Hispanic, and Asian mothers in the poorest ones, demonstrating a gap despite factors that should improve Black mothers' outcomes --- living in the same place, and having a higher income --- likely due to societal bias that impacts Black women \cite{hoffman2016racial}. 
\item Finally, after a model is trained, \textbf{post-deployment considerations} may not fully consider the impact of deploying a biased prediction model into clinical settings that have large Black populations. Because Black women have a heightened risk of pregnancy-related death across income and education levels~\cite{creanga2014maternal}, a biased prediction model could potentially automate policies or risk scores that disadvantage Black mothers.

\end{enumerate}
\begin{marginnote}[]
\entry{Model outcome}{The output of interest for predictive models}
\entry{Deployment}{The process through which a machine learning model is integrated into an existing production environment}
\entry{Risk score}{A calculated number denoting the likelihood of adverse event}
\end{marginnote}

We organize the rest of this review sequentially expanding on each of the five steps in the pipeline described above and in \textbf{Figure~\ref{fig:pipeline}}. First, we look at problem selection, and explain how funding for ML for health research can lead to injustice. We then examine how data collection processes in funded research can amplify inequity and unfairness. We follow this by exploring outcome definition and algorithm building, listing the multitude of factors that can impact model performance, and how these differences in performance relate to issues of justice. We close with audits that should be considered for more robust and just deployments of models in health, and recommendations to practitioners for ethical, fair, and just ML deployments. 

\section{PROBLEM SELECTION}
\label{chap29:sec2}
There are many factors that influence the selection of a research problem, from interest to available funding. However, problem selection can also be a matter of justice if the research questions that are proposed, and ultimately funded, focus on the health needs of advantaged groups. Below we provide examples of how disparities in research teams and funding priorities exacerbate existing socioeconomic, racial, and gender injustices. 

\subsection{Global Health Injustice} The ``10/90'' gap refers to the fact that the vast majority of health research dollars are spent on problems that affect a small fraction of the global population~\cite{vidyasagar2006global,leah2020health}.
Diseases that are most common in lower-income countries receive far less funding than diseases that are most common in high-income countries~\cite{von2015poverty} (relative to the number of individuals they affect). As an example, 26 poverty-related diseases account for 14\% of the global disease burden, but receive only 1.3\% of global health-related research and development expenditure. Nearly 60\% of the burden of poverty-related neglected diseases occurs in Western and Eastern sub-Saharan Africa as well as South Asia. Malaria, tuberculosis, and HIV/AIDS all have shares of global health-related research and development expenditure that are at least five times smaller than their share of global disease burden~\cite{von2015poverty}. This difference in rates of funding represents an injustice because it further exacerbates the disadvantages faced by Global South populations. While efforts like the ``All Of Us'' Project~\cite{all2019all} and the 23andMe's Call for Collaboration~\cite{under_2019} seek to collect more inclusive data, these efforts have come under criticism for not reflecting global health concerns, particularly among Indigenous groups~\cite{tsosie2019overvaluing}. %kaiser2019native,

\begin{marginnote}[]
\entry{Global South}{Countries on one side of the North–South divide, the other side being the countries of the Global North}
\end{marginnote}

\subsection{Racial Injustice} Racial bias affects which health problems are prioritized and funded. For example, sickle cell disease and cystic fibrosis are both genetic disorders of similar severity, but sickle cell disease is more common in Black patients, while cystic fibrosis is more common in white patients. In the United States (US), however, cystic fibrosis receives 3.4 times more funding per affected individual from the US National Institutes of Health (NIH), the largest funder of US clinical research, and hundreds of times more private funding~\cite{farooq2018disparities}. The disparities in funding persist despite the 1972 Sickle Cell Anemia Control Act, which recognizes that sickle cell has been neglected by the wider research community. Further, screening for sickle cell disease is viewed by some as unfair targeting~\cite{park_2010}, and Black patients with the disease who seek treatment are often maligned as drug abusers~\cite{rouse2009uncertain}.

\subsection{Gender Injustice} Women's health conditions like endometriosis are poorly understood; as a consequence, even basic statistics like the prevalence of endometriosis remain unknown, with estimates ranging from 1\% to 10\% of the population~\cite{chakradhar2018discovery,eisenberg2018epidemiology}. Similarly, the menstrual cycle is stigmatized and understudied~\cite{chakradhar2018discovery,pierson2019menstrual}, producing a dearth of understanding that undermines the health of half the global population. Basic facts about the menstrual cycle --- including which menstrual experiences are normal and which are predictive of pathology --- remain unknown~\cite{chakradhar2018discovery}. This lack of focus on the menstrual cycle propagates into clinical practice and data collection despite evidence that it affects many aspects of health and disease~\cite{hillard2014menstruation,american2006menstruation}. Menstrual cycles are also not often recorded in clinical records and global health data~\cite{chakradhar2018discovery}. In fact, the NIH did not have an R01 grant, the NIH's original and historically oldest grant mechanism, relating to the influence of sex and gender on health and disease until 2019~\cite{nih}. 
Notably, recent work has moved to target such understudied problems via ambulatory women's health-tracking mobile apps. These crowd-sourcing efforts stand to accelerate women's health research by collecting data from cohorts that are orders of magnitude larger than those used in previous studies~\cite{chakradhar2018discovery}. 

\subsection{Diversity of the Scientific Workforce}\label{subsec:ps_div}
The diversity of the scientific workforce profoundly influences the problems studied, and contributes to the biases in problem selection~\cite{kasy2020fairness}. Research shows that scientists from underrepresented racial and gender groups tend to prioritize different research topics. They produce more novel research, but their innovations are taken up at lower rates~\cite{hofstra2020diversity}. Female scientists tend to study different scientific subfields, even within the same larger field --- for example, within sociology, they have been historically better-represented on papers about sociology of the family or early childhood ~\cite{west2013role} --- and express different opinions about ethical dilemmas in computer science~\cite{pierson2017demographics}. Proposals from white researchers in the US are more likely to be funded by the NIH than proposals from Black researchers~\cite{hoppe2019topic,ginther2011race}, which in turns affects what topics are given preference. For example, a higher fraction of NIH proposals from Black scientists study community and population-level health~\cite{hoppe2019topic}. Overall, this evidence suggests that diversifying the scientific workforce will lead to problem selection that more equitably represents the interests and needs of the population as a whole.

\section{DATA COLLECTION}
\label{sec:data_collection}
The role of health data is ever-expanding, with new data sources routinely being integrated into decision-making around health policy and design. This wealth of high-quality data, coupled with advancements in ML models, has played a significant role in accelerating the use of computationally informed policy and practice to strengthen health care and delivery platforms. 
Unfortunately, data can be biased in ways that have (or can lead to) disproportionate negative impacts on already marginalized groups. 
First, data on group membership can be completely absent. For instance, countries such as Canada and France do not record race and ethnicity in their nationalized health databases~\cite{cihi,leonard2014census}, making it impossible to study race-based disparities and hypotheses around associations of social determinants of health. Second, data can be imbalanced. Recent work on acute kidney injury achieved state-of-the-art prediction performance in a large dataset of 703,782 adult patients using 620,000 features; however, they note that model performance was lower in female patients since female patients comprised 6.38\% of patients in the training data ~\cite{tomavsev2019clinically}. Other work has indicated that this issue can not be simply addressed by ``pre-training'' a model in a more balanced data setting prior to fine-tuning on an imbalanced dataset~\cite{mcdermott2020comprehensive}. This indicates that a model cannot be ``initialized'' with a balanced baseline representation which ameliorates issues of imbalance in downstream tasks, and suggests we must solve this problem at the root, be it with more balanced are comprehensive data, specialty learning algorithms, or combinations therein.%~\cite{seyyed2020chexclusion}
Finally, while some sampling biases can be recognized and possibly corrected, others may be difficult to correct. For example, work in medical imaging has demonstrated that models may overlook unforeseen stratification of conditions, like rare manifestations of diseases, which can result in harm in clinical settings~\cite{oakden2020hidden, ustun2019fairness}.

In this section, we discuss common biases in data collection. We consider two types of processes that result in a loss of data. First, processes that affect what kind of information is collected, or heterogenous data loss, across varying input types. For example, clinical trials with aggressive inclusion criteria or social media data that reflects those with access to devices. Second, we examine processes that affect whether an individual's  information is collected, or population-specific data losses, where individuals are impacted by their population type, often across data input categories. For example, undocumented immigrants may fear deportation if they participate in health care systems.

\subsection{Heterogeneous Data Losses}
Some data loss is specific to the data type, due to assumptions about noise that may have been present during the collection process. However, data noise and missingness can cause unjust inequities that impact populations in different ways. We cover four main data types: randomized controlled trials (RCTs), electronic health care records (EHR), administrative health data, and social media data. 

\textbf{Randomized Controlled Trials}
Randomized controlled trials are often run specifically to gather ``unbiased'' evidence of treatment effects. However, RCTs have notoriously aggressive exclusion (or inclusion) criteria~\cite{rothwell2005external}, which create study cohorts that are not representative of general patient populations~\cite{courtright2016point}. In one study of RCTs used to define asthma treatment, an estimated 94\% of the adult asthmatic population would not have been eligible for the trials~\cite{travers2007external}. 
There is a growing methodological literature designing methods to generalize RCT treatment effects to other populations~\cite{stuart2015assessing}. However, current empirical evidence indicates that such generalizations can be challenging given available data or may require strong assumptions in practice.

\begin{marginnote}[]
\entry{Training data}{Information that a ML model fits to and learns patterns from}
\entry{Heterogeneous data loss}{The process where data can be lost in collection due to data type}
\entry{Population-specific data loss}{The process where data can be lost in collection due to the features of the population}
\entry{Data noise}{Meaningless information added to data that obscures the underlying information of the data}
\entry{Missingness}{The manner in which data is absent from a sample of the population}
\entry{Randomized controlled trial (RCT)}{A study in which subjects are allocated by chance to receive one of several interventions}
\entry{Electronic health record (EHR)}{Digital version of a patient's clinical history that is maintained by the provider over time}
\entry{Intervention}{A treatment, procedure, or other action taken to prevent or treat disease, or improve health in other ways}
\end{marginnote}

\textbf{Electronic Health Records}
Much recent work in ML also leverages large electronic health records data. EHR data are a complex reflection of patient health, health care systems, and providers, where data missingness is a known, and meaningful, problem~\cite{wells2013strategies}. As one salient example, a large study of laboratory tests to model three-year survival found that health care process features had a stronger predictive value than the patient's physiological feaures~\cite{agniel2018biases}. Further, not all treatments investigated in RCTs can be easily approximated in EHR~\cite{bartlett2019feasibility}.

Biases in EHR data may arise due to differences in patient populations, access to care, or the availability of EHR systems~\cite{ferryman2018fairness}. As an example, the widely-used MIMIC-III EHR dataset includes most patients who receive care at the intensive care units in Beth Israel Deaconess Medical Center (BIDMC), but this sample is obviously limited by which individuals have access to care at BIDMC, which has a largely white patient population~\cite{chen2018my}. In the United States, uninsured Black and Hispanic or Latin(o/x) patients, as well as Hispanic or Latin(o/x) Medicaid patients, are less likely to have primary care providers with EHR systems, as compared to white patients with private insurance~\cite{hing2009there}.
Other work has shown that gender discrimination in health care access has not been systematically studied in India, primarily due to a lack of reliable data \cite{kapoor2019missing}. 

\textbf{Administrative Health Records}
In addition to RCTs and EHR, health care billing claims data, clinical registries, and linked health survey data are also common data sources in population health and health policy research~\cite{haneuse2017use,wing2018designing}, with known biases in which populations are followed, and who is able to participate. Translating such research into practice is a crucial part of maintaining health care quality, and limited participation of minority populations by sexual orientation and gender identity \cite{callahan2014eliminating}, race and ethnicity \cite{lopez2017discrepancies}, and language \cite{klinger2015accuracy} can lead to health interventions and policies that are not inclusive, and can create new injustices for these already marginalized groups.

\textbf{Social Media Data}
Data from social media platforms and search-based research by nature consists only of individuals with internet access~\cite{dredze2012social}. Even small choices like limiting samples to those from desktop versus mobile platforms are a problematic distinction in non-North American contexts \cite{abebe2019using}.
Beyond concerns about access to resources or geographic limitations, data collection and scraping pipelines for most social media cohorts do not yield a random sample of individuals. 
Further, the common practice of limiting analysis to those satisfying a specified threshold of occurrence can lead to skewed data. As an example, when processing the large volume of Twitter data (7.6 billion tweets) researchers may first restrict to users who can be mapped to a US county (1.78 billion), then to those Tweets that contain only English (1.64 billion tweets), and finally remove users who made less than 30 posts (1.53 billion) \cite{giorgi2018remarkable}.

\subsection{Population-specific Data Losses}
As with data types, the modern data deluge does not apply equally to all communities. Historically underserved groups are often underrepresented, misrepresented, or entirely missing from health data that inform consequential health policy decisions. When individuals from disadvantaged communities appear in observational datasets, they are less likely to be accurately captured due to errors in data collection and systemic discrimination. Larger genomics datasets often target European populations, producing genetic risk scores which are more accurate in individuals of European ancestry than other ancestries~\cite{martin2019clinical}. We note four specific examples of populations that are commonly impacted: low- and middle-income nationals, transgender and gender non-conforming individuals, undocumented migrants, and pregnant women. 

\textbf{Low- and Middle-Income Nationals}
Health data are infrequently collected due to resource constraints, and even basic disease statistic data such as prevalence of mortality rates can be challenging to find for low- and middle-income nations \cite{abebe2019using}. When data are collected, it is not digitized, and often contains errors. In 2001, the World Health Organization found that only 9 out of the 46 member states in Sub-Saharan Africa could produce death statistics for a global assessment of the burden of disease, with data coverage often less than 60\% in these countries~\cite{jamison2006disease}.

\textbf{Transgender and Gender Non-conforming Individuals}
The health care needs and experiences of transgender and gender non-conforming individuals are not well-documented in datasets \cite{james2016report} because documented sex, not gender identity, is what is usually available. However, documented sex is often discordant with gender identity for transgender and gender non-conforming individuals. Apart from health documentation concerns, transgender people are often concerned about their basic physical safety when reporting their identities. In the US, it was only in 2016, with the release of the US Transgender Survey that there was a meaningfully sized dataset --- 28,000 respondents --- to enable significant analysis and quantification of discrimination and violence that transgender people face \cite{james2016report}. 

\textbf{Undocumented Immigrants}
Safety concerns are important in data collection for undocumented migrants, where socio-political environments can lead to individuals feeling unsafe during reporting opportunities. When immigration policies limit access to public services for immigrants and their families, these restrictions lead to spillover effects on clinical diagnoses. As one salient example, autism diagnoses for Hispanic children in California fell following aggressive federal anti-immigrant policies requiring citizenship verification at hospitals~\cite{fountain2011risk}. 

\textbf{Pregnant Women}
Despite pregnancy being neither rare nor an illness, the US continues to experience rising maternal mortality and morbidity rates. In the US, the maternal mortality rate has more than doubled from 9.8 per 100,000 live births in 2000 to 21.5 in 2014~\cite{ai2019maternal}.
Importantly, disclosure protocols recommend suppression of information in nationally available datasets when the number of cases or events in a data ``cell'' is low, to reduce the likelihood of a breach of confidentiality. For example, the US Centers for Disease Control suppresses numbers for counties with fewer than 10 deaths for a given disease~\cite{tiwari2014impact}. Although these data omissions occur because of patient privacy, such censoring on the \emph{dependent} variable introduces particularly pernicious statistical bias and, as a result, much remains to be understood about what community, health facility, patient, and provider-level factors drive high mortality rates. 

\begin{marginnote}[]
\entry{Censoring}{The mechanism through which data values are removed from observation}
\end{marginnote}

\section{OUTCOME DEFINITION}
\label{chap29:sec3}
The next step in the model pipeline is to define the outcome of interest for a health care task. Even seemingly straightforward tasks like defining whether a patient \emph{has} a disease can be skewed by how prevalent diseases are, or how they manifest in some patient populations. For example, a model predicting if a patient will develop heart failure will need labeled examples both of patients who have heart failure, and patients without heart failure. Choosing these patients can rely on parts of the EHR that may be skewed due to lack of access to care, or abnormalities in clinical care: e.g., economic incentives may alter diagnosis code logging~\cite{kesselheim2005overbilling}, clinical protocol affects the frequency and observation of abnormal tests~\cite{agniel2018biases}, historical racial mistrust may delay care and affect patient outcomes~\cite{boag2018racial}, and naive data collection can yield inconsistent labels in chest X-rays~\cite{oakden2020hidden}. Such biased labels, and the resulting models, may cause clinical practitioners to allocate resources poorly.

We discuss social justice considerations in two examples of commonly modelled health care outcomes: clinical diagnosis and health care costs. In each example, it is essential that model developers choose a reliable proxy and account for noise in the outcome labels as these choices can have a large impact on performance and equity of the resulting model.

\subsection{Clinical Diagnosis}
Clinical diagnosis is a fundamental task for clinical prediction models, e.g., models for computer-aided diagnosis from medical imaging. In clinical settings, researchers often select patient disease occurrence as the prediction label for models. However, there are many options for the choice of a disease occurrence label.
For example, the outcome label for developing cardiovascular disease could be defined through the occurrence of specific phrases in clinical notes. However, women can manifest symptoms of acute coronary syndrome differently~\cite{canto2007symptom} and receive delayed care as a result~\cite{bugiardini2017delayed},
which may then manifest in diagnosis labels derived from the clinical notes being gender-skewed. Because differences in label noise results in disparities in model impact, researchers have the responsibility to choose and improve disease labels, so that these inequalities do not further exacerbate disparities in health. 

Additionally, it is important to consider the health care system in which disease labels are logged. For example, health care providers leverage diagnosis codes for billing purposes, not clinical research. As a result, diagnosis codes can create ambiguities because of overlap and hierarchy in codes. Moreover, facilities have incentives to under-report~\cite{kesselheim2005overbilling} and over-report~\cite{rose2016machine,geruso2015upcoding} outcomes, yielding differences in model representations. 

\begin{marginnote}[]
\entry{Label noise}{Errors or otherwise obscuring information that
affects the quality of the labels}
\entry{Diagnosis code}{A label in patient records of disease occurrence, which may be subject to misclassification, used primarily for billing purposes}
\end{marginnote}

Recent advances in improving disease labels target statistical corrections based on estimates of the label noise. For instance, a positive label may be reliable, but the omission of a positive label could either indicate a negative label (i.e., no disease) or merely a missed positive label. Methods to address the positive-unlabeled setting use estimated noise rates~\cite{natarajan2013learning} or hand-curated labels that are strongly correlated with positive labels, known also as ``silver-standard'' labels, from clinicians~\cite{halpern2016electronic}. Clinical analysis of sources of error in disease labels can also guide improvements~\cite{oakden2020exploring} and identify affected groups~\cite{oakden2020hidden}.

\subsection{Health Care Costs}
Developers of clinical models may choose to predict health care costs, meaning the ML model seeks to predict which patients will cost the health care provider more in the future. Some model developers may use health care costs as a proxy for future health needs to guide accurate targeting of interventions~\cite{obermeyer2019dissecting}, with the underlying assumption that addressing patients with future health need will limit future costs. Others may explicitly want to understand patients who will have high health care cost to reduce the total cost of health care~\cite{tamang2017predicting}.
However, because socioeconomic factors affect both access to health care \emph{and} access to financial resources, these models may yield predictions that exacerbate inequities.  

For model developers seeking to optimize for health need, health care costs can deviate from health need on an individual level because of patient socioeconomic factors. For instance, in a model used to allocate care management program slots to high-risk patients, 
the choice of future health care costs as a predictive outcome led to racial disparities in patient allocation to the program~\cite{obermeyer2019dissecting}.
Health care costs can differ from health need on an institutional level due to underinsurance and undertreatment within the patient population~\cite{cook2012measuring}. After defining health disparities as all differences except those due to clinical need and preferences, researchers have found racial disparities in mental health care. Specifically, white patients had higher rates of initiation of treatment for mental health compared to Black and Hispanic or Latin(o/x) patients. Because the analysis controls for health need, the disparities are solely a result of differences in health care access and systemic discrimination~\cite{cook2014assessing}.

Addressing issues that arise from the use of health care costs depends on the setting of the ML model. In cases where health need is of highest importance, a natural solution is to choose another outcome definition besides health care costs, e.g., the number of chronic diseases as a measure of health need. If a model developer is most concerned with cost, it is possible to correct for health disparities in predicting health care costs by building fairness considerations directly into the predictive model objective function~\cite{zink2019fair}. Building these types of algorithmic procedures is further discussed in Section~\ref{sec:building_algorithm}.

\section{ALGORITHM DEVELOPMENT}
\label{sec:building_algorithm}
Algorithm development considers the construction of the underlying computation for the ML model and presents a major vulnerability and opportunity for ethical ML in health care. Just as data are not neutral, algorithms are not neutral. 
A disproportionate amount of power lies with research teams who, after determining the research questions, make decisions about critical components of an algorithm such as the loss function~\cite{guillory2020combating,kasy2020fairness}. 
In the case of loss functions, common choices like the $L_1$ absolute error loss and $L_2$ squared error loss do not target the same conditional functions of the outcome, instead minimizing the error in the median and mean respectively.
\begin{marginnote}[]
\entry{Loss function}{The relation that determines the error between algorithm output and given label, which the algorithm uses to optimize}
\end{marginnote}
Using a surrogate loss (e.g., hinge loss for the error rate) can provide computational efficiency, but it may not reflect the ethical criteria that we truly care about. Recent work has shown that models trained with a surrogate loss may exhibit ``approximation errors'' that disproportionately affect undersampled groups in the training data~\cite{lohaus2020too}. Similarly, one might choose to optimize the worst-case error across groups as opposed to the average overall error. Such choices may seem purely technical, but reflect value statements about what should be optimized, potentially leading to differences in performance among marginalized groups~\cite{sagawa2019distributionally}. 

In this section, we review several crucial factors in model development that potentially impact ethical deployment capacity: understanding (and accounting for) confounding, feature selection, tuning parameters, and defining ``fairness'' itself. 

\subsection{Understanding Confounding}
Developing models that use sensitive attributes without a clear causal understanding of their relationship to outcomes of interest can significantly affect model performance and interpretation. This is relevant to algorithmic problems focused on prediction, not just causal inference. ``Confounding'' features --- i.e., those features that influence both the independent variables and the dependent variable --- require careful attention. 
The vast majority of models learn patterns based on observed correlations between training data, even when such correlations do not occur in test data. For instance, recent work has demonstrated that classification models designed to detect hair color learn gender-biased decision boundaries when trained on confounded data, i.e., if women are primarily blond in training data, the model incorrectly associates gender with the hair label in test samples ~\cite{joshi2018xgems}. 

\begin{marginnote}[]
\entry{Confounding}{The condition in which a feature influences both the dependent variable and independent variable, causing a spurious association}
\entry{Test data}{Unseen information that a model predicts on and is evaluated against}
\end{marginnote}

As ML methods are increasingly used for clinical decision support, it is critical to account for confounding features. In one canonical example, asthmatic patients presenting with pneumonia are given aggressive interventions that ultimately improve their chances of survival over non-asthmatic patients~\cite{caruana2015intelligible}. When the hospital protocol assigned additional treatment to patients with asthma, those patients had improved outcomes. Thus the treatment policy was a confounding factor that altered the data in a seemingly straight-forward prediction task such that patients with asthma were erroneously predicted by models to have \emph{lower} risk of dying from pneumonia. 

Simply controlling for confounding features by including them as features in classification or regression models may be insufficient to learn reliable models because features can have a mediating or moderating effect (post-treatment effect on outcomes of interest) and have to be incorporated differently into model design~\cite{hernan2010causal}.

Modern ML and causal discovery techniques can identify sources of confounding at scale~\cite{glymour2019review}, although validation of such methods can be challenging because of the lack of counterfactual data. ML methods have also been proposed to estimate causal effects from observational data~\cite{van2011targeted,chernozhukov2018double}.
In practice, when potential hidden confounding is suspected, either mediating features or proxies can be leveraged~\cite{miao2018identifying,hernan2010causal} or sensitivity analysis methods can be used to determine potential sources of errors in effect estimates~\cite{franks2019flexible}.
Data-augmentation and sampling methods may also be used to mitigate effects of model confounding. For example, augmenting X-ray images with rotated and translated variants can help train a model that is not sensitive to orientation of an image~\cite{little2019causal}. 

\subsection{Feature Selection}
With large-scale digitization of EHR and other sources, sensitive attributes like race and ethnicity may be increasingly available (although prone to misclassification and missingness). However, blindly incorporating factors like race and gender in a predictive model may exacerbate inequities for a wide range of diagnostics and treatments~\cite{vyas2020hidden}. These resulting inequities can lead to unintended and permanent embedding of biases in algorithms used for clinical care. For example, vaginal birth after cesarean (VBAC) scores are used to predict success of ``trial of labor" of pregnant women with a prior cesarean section; however, these scores explicitly include a race component as an input which reduces the chance of VBAC success for Black and Hispanic women. Although researchers found that previous observational studies showed correlation between racial identity and success of trial of labor~\cite{grobman2007development}, the underlying cause of this association is not well-understood. Such na\"ive inclusion of race information could exacerbate disparities in maternal mortality. This ambiguity calls race-based `correction' in scores like VBAC into question~\cite{vyas2020hidden}.

\begin{marginnote}[]
\entry{Sensitive attribute}{A specified patient feature (e.g., race, gender) which is considered important for fairness considerations}
\entry{Stepwise regression}{A method of estimation where each feature is sequentially considered by addition or subtraction to the existing feature set}
\end{marginnote}

Automation in feature selection does not eliminate the need for contextual understanding. For example, stepwise regression is commonly used and taught as a technique for feature selection despite known limitations~\cite{thompson1995stepwise}. 
While specific methods have varying initialization (e.g., start with an empty set of features or full set of features) and processing steps (e.g., deletion vs. addition of features), most rely on $p$-values, $R^2$, or other global fit metrics to select features. Weaknesses of stepwise regressions include the misleading nature of $p$-values and the final set depending on if and when features were considered~\cite{harrell2015regression}. In ML, penalized regressions like lasso regression are popular for automated feature selection, but the lasso trades potential increases in estimation bias for reductions in variance by shrinking some feature coefficients to zero. Features selected by lasso may be co-linear with other features not selected~\cite{james2013introduction}. Over-interpretation of the selected features in any automated procedures should therefore be avoided in practice given these pitfalls. Researchers should also consider the humans-in-the-loop framework where incorporation of automated procedures is blended with investigator knowledge~\cite{koh2020concept}. 

\subsection{Tuning Parameters}
There are many tuning parameters that may be set a priori or selected via cross-validation~\cite{james2013introduction}. These range from the learning rate in a neural network to the minimum size of the terminal leaves in a random forest. In the latter example, default settings in R for classification will allow trees to grow until there is just one observation in a terminal leaf. This can lead to overfitting the model to the training data and a loss of generalizability to the target population. Lack of generalizability is a central concern for ethical ML given the previously discussed issues in data collection and study inclusion. When data lack diversity and are not representative of the target population where the model would be deployed, overfitting algorithms to this data has the potential to disproportionately harm marginalized groups~\cite{sagawa2020investigation}. Using cross-validation to select tuning parameters does not automatically solve these problems as cross-validation still operates with respect to an a priori chosen optimization target. 

\begin{marginnote}[]
    \entry{Tuning parameters}{Algorithm components used for prediction that are tuned toward solving an optimization problem}
    \entry{AUC}{A measure of the sensitivity and specificity of a model for each decision threshold}
    \entry{AUPRC}{A measure of precision and recall of a model for each decision threshold}
    \entry{Calibration}{A measure of how well ML risk estimates reflect true risk}
\label{margin:1}
\end{marginnote}

\subsection{Performance Metrics} 
There are many commonly used performance metrics for model evaluation such as area under the receiver-operating characteristic (AUC), precision-recall curves (AUPRC), and calibration~\cite{flach2003geometry}. However, the appropriate metrics to optimize depend on intended use case and relative value of true positives, false positives, true negatives, and false negatives. 
Not only can AUC be misleading when considering other global fit metrics (e.g., high AUC masking weak true positive rate), but it does not describe the impact of the model across selected groups. Further, even ``objective'' metrics and scores can be deeply flawed, and lead to over or under-treatment of minorities if blindly applied \cite{hidden2020NEJM}.
Note that robust reporting of results should include an explicit statement of other non-optimized metrics, including the original intended use case, the training cohort and case, or level of model uncertainty. 

\subsection{Group Fairness Definition} 
The specific definition of fairness for a given application often impacts the choice of a loss function, and therefore the underlying algorithm. Individual fairness imposes classifier performance requirements that operate over pairs of individuals, e.g., similar individuals should be treated similarly~\cite{dwork2012fairness}. Group fairness operates over ``protected groups'' (based on some sensitive attribute) by requiring that a classifier performance metric be balanced across those groups. ~\cite{dwork2018group,chouldechova2020snapshot}. For instance, a model may be partially assessed by calculating the true positive rate separately among rural and urban populations to ensure risk score similarity. 
Regressions subject to group fairness constraints or penalties optimizing toward joint global and group fit considerations have also been developed~\cite{calders2013controlling,zafar2017fairnessa,zink2019fair}.  
\begin{marginnote}[]
    \entry{Performance metric}{Score or other quantitative representation of a model's quality and ability to achieve goals}
\entry{Algorithmic fairness}{The study of definitions and methods related to the justice of models}
\entry{Group fairness}{A principle where pre-defined patient groups should receive similar model performance}
\end{marginnote}

Recent work has focused on identifying and mitigating violations of fairness definitions in health care settings. While most of these algorithms have emerged outside the field of health care, researchers have designed penalized and constrained regressions to improve the performance of health insurance plan payment. This payment system impacts tens of millions of lives in the United States and is known to undercompensate insurers for individuals with certain health conditions, including mental health and substance use disorders, in part because billing codes do not accurately capture diagnoses~\cite{montz2016risk}. Undercompensation creates incentives for insurers exclude individuals with these health conditions from enrollment, limiting their access to care. Regressions subject to group fairness constraints or penalties were successful in removing nearly all undercompensation for a single group with negligible impacts on global fit~\cite{zink2019fair}. Subsequent work incorporating multiple groups into the loss function also saw improvements in undercompensation for the majority of groups not included ~\cite{mcguire2020simplifying}.

\section{POST-DEPLOYMENT CONSIDERATIONS}
\label{sec:post_deployment}
Often the goal of model training is to ultimately deploy it in a clinical, epidemiological, or policy service. However, deployed models can have lasting ethical impact beyond the model performance measured in development. For example, in the inclusion of race in the clinical risk scores described earlier that may lead to chronic over- or under-treatment~\cite{vyas2020hidden}. Here we outline considerations for robust deployment by highlighting the need for careful performance reporting, auditing generalizability, documentation, and regulation.

\subsection{Quantifying Impact}
Unlike in other settings with high-stakes decisions, e.g., aviation, clinical staff performance is not audited by an external body \cite{helmreich2000error}. Instead, clinicians are often a self-governing body, relying on clinicians themselves to determine when a colleague is underperforming or in breach of ethical practice principles, e.g., through such tools as surgical morbidity and mortality conferences \cite{murayama2002critical}. Clinical staff can also struggle to keep abreast of what current best practice recommendations are, as these can change dramatically over time; one study found that more than 400 previously routine practices were later contradicted in leading clinical journals \cite{herrera2019meta}.

Hence, it is important to measure and address the downstream impact of models though audits for bias and examination of clinical impact~\cite{chen2019can}.
Regular ``auditing'' post-deployment, i.e., detailed inspection of model performance on various groups and outcomes, may reveal the impact of models on different populations~\cite{obermeyer2019dissecting} and identify areas of potential concern. Some recent work has targeted causal models in dynamic systems in order to reduce the severity of bias \cite{creager2019causal}. Others have targeted bias reduction through model construction with explicit guarantees about balanced performance \cite{ustun2019fairness}, or specifying groups which must have equal performance \cite{noseworthy2020assessing}. Additionally, there is the possibility that models may help to de-bias current clinical care by reducing known biases against minorities \cite{treatment2002confronting} and disadvantaged majorities \cite{perez2019invisible}.

\begin{marginnote}[]
\entry{Model auditing}{The post-deployment inspection of model performance on groups and outcomes}
\end{marginnote}

\subsection{Model Generalizability}
\label{sec:model_generalizability}
As has been raised in previous sections, a crucial concern with model deployment is generalization. Any shifts in data distributions can significantly impact model performance when the settings for development and for deployment differ. For example, chest X-ray diagnosis models can have high performance on test data drawn from the same hospital but degrade rapidly on data from another hospital~\cite{zech2018variable}. Other work in gender bias on chest X-ray data has demonstrated both that small proportions of female chest x-rays degrades diagnostic performance accuracy in female patients \cite{larrazabal2020gender}, and that this is not simply addressed in all cases by adding in more female X-rays \cite{seyyed2020chexclusion}. 
Even within a single hospital, models trained on data from an initial EHR system data deteriorated significantly when tested on data from a new EHR system ~\cite{nestor2019feature}. Finally, data artifacts that induce strong priors in what patterns ML models are sensitive to have the potential to perpetrate harms when used without awareness~\cite{bissoto2019constructing}. For example, patients with dark skin can have morphological variation and disease manifestations that are not easily detected under the defaults that are set by predominantly white-skinned patients~\cite{kundu2013dermatologic}. 

\begin{marginnote}[]
\entry{Generalizability}{The ability of a model to apply in a setting different from the one in which it was trained}
\entry{Data artifact}{A flaw in data caused by equipment, techniques, or conditions that is unrelated to model output}
\end{marginnote}

Several algorithms have recently been proposed to account for distribution shift in data~\cite{NIPS2019_8420,subbaswamy2020development}. However, these algorithms have significant limitations as they typically require assumptions about the nature or amount of distributional shift an algorithm can accommodate. Some, like~\cite{subbaswamy2020development}, may require a clear indication of which distributions in a health care pipeline are expected to change, and develop models for prediction accordingly. Many of these assumptions may be verifiable. If not, periodically monitoring for data shifts~\cite{davis2017calibration}, and potentially retraining models when performance deteriorates due to such shifts is an imperative deployment consideration with significant ethical implications.

\subsection{Model and Data Documentation}
Clear documentation enables insight into the model development and data collection. Good model documentation should include clinically specific features of model development that can be assessed and recorded beforehand, such as logistics within the clinical setting, potential unintended consequences, and trade-offs between bias and performance~\cite{salehclinical}. In addition to raising ethical concerns in the pipeline, the process of co-designing ``checklists" with clinical practitioners formalizes ad-hoc procedures and empowers individual advocates~\cite{madaio2020co}. Standardized reporting of model performance---such as the one-page summary ``model cards'' for model reporting~\cite{mitchell2019model}---can empower clinical practitioners to understand model limitations and future model developers to identify areas of improvement. Similarly, better documentation of the data supporting initial model training can help expose sources of discrimination in the collected data. Modelers could use ``datasheets'' for datasets to detail the conditions of data collection~\cite{gebru2018datasheets}. 

\subsection{Regulation}
In the United States, the Food and Drug Administration (FDA) has responsibility for the regulation of health care ML models. As there does not exist comprehensive guidance for health care model research and subsequent deployment, the opportunity is ripe to create a comprehensive framework to audit and regulate models. Currently, the FDA's proposed ML-specific modifications to the software as a medical device (SaMD) regulations draw a distinction between models that are trained and then frozen prior to clinical deployment and models that continue to learn on observed outcomes. Although models in the latter class can leverage larger, updated datasets, they also face additional risk due to model drift and may need additional audits~\cite{cdc2020}.
Such frameworks should explicitly account for health disparities across the stages of ML development in health, and ensure health equity audits as part of postmarket evaluation~\cite{ferryman2015addressing}. We also note that there are many potential legal implications, e.g., in malpractice and liability suits, that will require new solutions~\cite{sullivan2019current}.

Researchers have proposed additional frameworks to guide clinical model development, which could inspire future regulation. ML model regulation could draw from existing regulatory frameworks: a randomized controlled trial for ML models would assess patient benefit compared to a control cohort of standard clinical practice~\cite{liu2019reporting}, and a drug development pipeline for ML models would define a protocol for adverse events and model recalls~\cite{coravos2019we}. The clinical interventions accompanying the clinical ML model should be analyzed to contextualize the use of the model in the clinical setting~\cite{parikh2019regulation}.

\section{RECOMMENDATIONS}

\begin{figure}
    \centering
    \includegraphics[width=0.95 \textwidth]{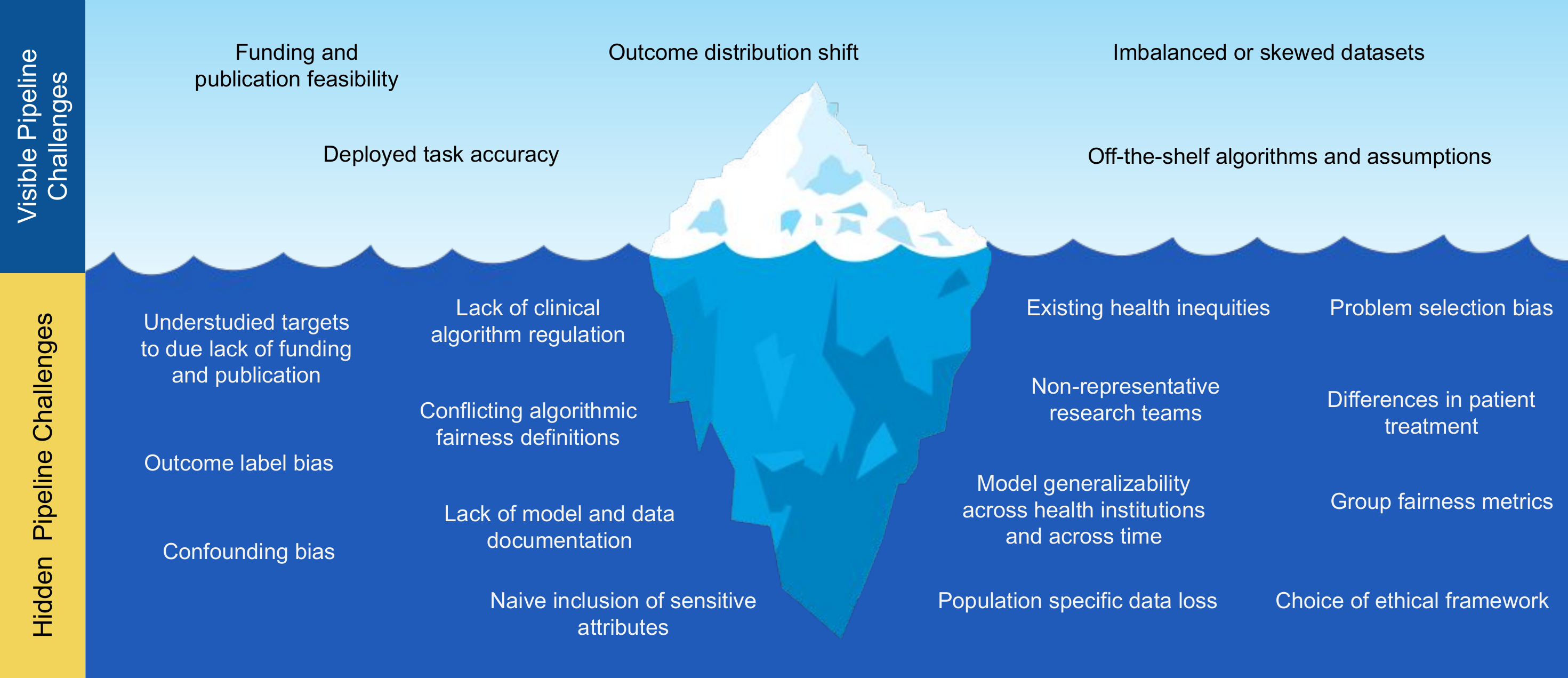}
    \caption{The model development pipeline contains many challenges for ethical machine learning for health care. We highlight both visible and hidden challenges.}
    \label{fig:iceberg}
\end{figure}

In this review, we have described the ethical considerations at each step of the ML model development pipeline we introduced. While most researchers will address known challenges like deployed task accuracy and outcome distribution shift, they are unlikely to be aware of the full magnitude of the hidden challenges such as existing health inequities or outcome label bias. 
As seen in \textbf{Figure~\ref{fig:iceberg}}, many hidden pipeline challenges can go unaddressed in a typical ML health project, but have serious ethical repercussions. With these challenges in mind, we propose five general recommendations that span the pipeline stages. 
\begin{enumerate}
    \item Problems should be tackled by diverse teams and using frameworks that increase the probability that equity will be achieved. Further, historically understudied problems are important targets to practitioners looking to perform high-impact work. 
    \item Data collection should be framed as an important front-of-mind concern in the ML modelling pipeline including clear disclosures about imbalanced datasets, and researchers should engage with domain experts to ensure that data reflecting underserved and understudied populations' needs are gathered. 
    \item Outcome choice should reflect the task at hand and should preferably be unbiased. If the outcome label has ethical bias, the source of inequity should be accounted for in ML model design, leveraging literature that attempts to remove ethical biases during pre-processing, or with use of a reasonable proxy. 
    \item Reflection on the goals of the model is essential during development, and should be articulated in a pre-analysis plan. In addition to technical choices like loss function, researchers must interrogate how, and whether, a model should be developed to best answer a research question, and what caveats are included. 
    \item Audits should be designed to identify specific harms, and paired with methods and procedures. Harms should be examined group-by-group, rather than at a population level. ML ethical design ``checklists'' are one possible tool to systematically enumerate and consider such ethical concerns prior to declaring success in a project. 
\end{enumerate}

Finally, we note that machine learning also could and should be harnessed to create shifts in power in health care systems~\cite{mohamed2020decolonial}. This might mean actively selecting problems for the benefit of underserved patients, designing methods to target systemic interventions for improved access to care and treatments, or enforcing evaluations with the explicit purpose of preserving patient autonomy.
In one salient example, the state of California reduced disparities in rates of obstetric hemorrhage (and therefore maternal mortality for women of color) by weighing blood loss sponges, i.e., making access to treatment consistent and unbiased for all women~\cite{lyndon2015cumulative}. Models could similarly be harnessed to learn and recommend consistent rules, giving researchers a potential opportunity to de-bias current clinical care~\cite{chen2020treating}, measure racial disparities and mistrust in end-of-life care~\cite{boag2018racial}, improve known biases against minorities \cite{treatment2002confronting} and disadvantaged majorities \cite{perez2019invisible}. Ultimately, the responsibility for ethical models and behavior lies with a broad community, but begins with technical researchers fulfilling an obligation to engage with patients, clinical researchers, staff, and advocates to build ethical models.

\begin{issues}[FUTURE QUESTIONS]
\begin{enumerate}
\item How can we combat urgent global health crises that exacerbate existing patterns of health injustices?
\item How can we encourage model developers to build ethical considerations into the pipeline from the very beginning? Currently, when egregious cases of injustice are discovered only after clinical impact has already occurred, what can developers do to engage? 
\item How can evaluation and audits of ML systems be translated into meaningful clinical practice when, in many countries, clinicians themselves are subject to only limited external evaluations or audits?
\item When, if ever, should sensitive attributes like race be used in analysis? How should we incorporate socially constructed features into models and audits?
\item How can ML be used to shift power, e.g., from well-known institutions, privileged patients, and wealthy multinational corporations to the patients most in need?
\end{enumerate}
\end{issues}

\section*{DISCLOSURE STATEMENT}
The authors are not aware of any affiliations, memberships, funding, or financial holdings that
might be perceived as affecting the objectivity of this review. 

% Acknowledgements
\section*{ACKNOWLEDGMENTS}

The authors thank Rediet Abebe for helpful discussions and contributions to an early draft, and Peter Szolovits, Pang Wei Koh, Leah Pierson, Berk Ustun, and Tristan Naumann for useful comments and feedback. This work was supported in part by an NIH Director's New Innovator Award DP2‐MD012722 (SR), a CIFAR AI Chair at the Vector Institute (MG), and Microsoft Research (MG).

\bibliographystyle{ieeetr}
\bibliography{reference}

\begin{thebibliography}{100}
\bibitem{nih}
``Nih offers its first research project grant (r01) on sex and gender.''
\end{thebibliography}

\end{document}